\documentclass{iaus}
\usepackage{graphicx}

\newcommand{\gtapprox}{\raisebox{-0.5ex}{$\,\stackrel{>}{\scriptstyle
\sim}\,$}}
\newcommand{\ltapprox}{\raisebox{-0.5ex}{$\,\stackrel{<}{\scriptstyle
\sim}\,$}}
\newcommand{\ri}{r_{\rm i}}

\newcommand{\LEdd}{L_{\rm Edd}}

\newcommand{\Lx}{L_{0.5-8 \,{\rm keV}}}

\newcommand{\Mdota}{\dot M_{\rm a}}

\newcommand{\Mbh}{M_{\bullet}}
\newcommand{\LX}{L_{\scriptscriptstyle \rm X}}
\newcommand{\NH}{N{\scriptscriptstyle \rm H}}

\title[ULXs: XRBs in a High/Hard State?] %% give here short title %%
{Ultra-Luminous X-ray Sources: X-ray Binaries in a High/Hard State?}

\author[Kuncic et al.]   %% give here short author list %%
{Z. Kuncic$^1$, %
 R. Soria$^2$, C.K. Hung$^1$, M.C. Freeland$^1$ \break \and G.V. Bicknell$^3$}

\affiliation{$^1$School of Physics, University of Sydney, Sydney NSW
  2006, Australia \break email: z.kuncic@physics.usyd.edu.au\\[\affilskip]
$^2$Mullard Space Science Laboratory, University College London,
  Holmbury St Mary, \break Dorking, Surrey RH5 6NT, UK \break email:
  rs1@mssl.ucl.ac.uk\\[\affilskip]$^3$Research School of Astronomy
and Astrophysics, Australian National University, \break Mount Stromlo
Observatory, Cotter Road, Canberra ACT 2611, Australia \break email: geoff@mso.anu.edu.au}

\pubyear{2006}
\volume{238}  %% insert here IAU Symposium No.
\pagerange{001--999}
\date{??? and in revised form ???}
\jname{Black Holes: from Stars to Galaxies -- across the Range of Masses}
\editors{V. Karas \& G. Matt, eds.}
\begin{document}

\maketitle

\begin{abstract}
We examine the possibility that Ultraluminous X-ray sources (ULXs)
represent the extreme end of the black hole X-ray binary (XRB)
population. Based on their X-ray properties, we suggest that ULXs are
persistently in a high/hard spectral state and we propose a new
disk--jet model that can accomodate both a high accretion rate and
a hard X-ray spectrum. Our model predicts that the modified disk
emission can be substantially softer than that predicted by a standard
disk as a result of jet cooling and this may explain the unusually
soft components that are sometimes present in the spectra of bright
ULXs. We also show that relativistic beaming of jet emission can
indeed account for the high X-ray luminosities of ULXs, but strong
beaming produces hard X-ray spectra that are inconsistent with
observations. We predict the beamed synchrotron radio emission should
have a flat spectrum with a flux density $\ltapprox 0.01\,{\rm mJy}$.
\end{abstract}

\firstsection % if your document starts with a section,
\section{Introduction}\label{sec:intro}

The exceptionally high X-ray luminosity of ULXs, $\Lx \approx (0.2 -
10) \times 10^{40}{\rm erg \,  s^{-1}}$, emerges almost entirely as a
hard power-law with photon indices $1.5 \ltapprox \Gamma
\ltapprox 2.5$ (\cite[Irwin, Bregman \& Athey
  2004]{Irwin04}; \cite[Swartz \etal\ 2004]{Swartz04}; \cite[Liu \&
  Bregman 2005]{LiuBreg05}; \cite[Stobbart, Roberts
  \& Wilms 2006]{Stobbart06}). ULXs are statistically consistent with
representing a sub-population of XRBs (\cite[Swartz
  \etal\ 2004]{Swartz04}). However, known
Galactic black hole XRBs have never been observed to enter a spectral
state in which they simultaneously have an X-ray luminosity that is at
least Eddington and a power-law spectrum that is harder than $\Gamma
\approx 2.5$. The XRB spectral state that perhaps most closely resembles ULXs
is the short-lived steep power-law (or very high) state, in which $\LX \approx
\LEdd$, but $\Gamma \gtapprox 2.5$ (\cite[see McClintock \& Remillard
  2006]{McClinRem06}). ULXs therefore appear to be in
a persistently high/hard state.

It is unclear whether existing models for XRBs can explain a high/hard
spectral state. ULXs appear to require a model with both a high
$\Mdota$ and strong corona and/or jet. Furthermore, it is clear that
most of the accretion power in ULXs is not being dissipated in a standard accretion disk. This implies that we cannot use the standard Multi-Colour Disk (MCD) model 
to fit any soft spectral component that might be present.  We
summarize the main results of the model proposed by \cite[Freeland
  \etal\ (2006)]{Freeland06} and we examine further the implications
of a strong, accretion-powered jet for spectral fitting of accretion disk models to soft components evident in some ULX spectra.

\section{Model Outline and Results}\label{sec:model}

The details of our ULX model can be found in \cite[Freeland
  \etal\ (2006)]{Freeland06}. It is based on the generalized
  theoretical framework of \cite{KunBick04}, which specifically
  addresses vertical angular momentum transport by a magnetic torque
  on the the accretion disk surface. This magnetic torque is
  identified as the mechanism responsible for jet/corona formation. It
  results in a modified radial structure of the accretion disk and
  hence, a modified disk spectrum, which we refer to as an Outflow
  Modified Multi-Colour Disk (OMMCD). Let us assume here that the  total
  power extracted by the torque is injected
  into a relativistic jet and partitioned into magnetic and kinetic energies. 

\subsection{Modified disk spectrum}

The radiative energy flux of an accretion disk modified by a
magnetized jet is
\begin{equation}
\sigma T^4 = F(r) = \frac{3G\Mbh \Mdota}{8\pi r^3} \left[ f_{\rm ss} (r) - f_{\rm
    j}(r) \right]
\end{equation} 
where $f_{\rm ss} = 1 - (r/r_{\rm i})^{-1/2}$ is the small-$r$
correction factor for a standard disk and where the jet correction
factor is directly related to the magnetic torque acting on the disk surface:
\begin{equation}
f_{\rm j}(r) = \frac{1}{\Mdota r^2 \Omega}\int_{\ri}^r 4\pi r^2
\frac{B_\phi B_z}{4\pi} \, {\rm d}r
\end{equation}
Thus, the jet drains energy from the disk and modifies the radial
temperature profile. As a result, the total disk spectrum is modified
in the soft X-ray bandpass for stellar-mass black holes (see also
\cite[Soria, Kuncic \& Gon\c{c}alves  2006]{SKG06}, this volume). This
means that unusually soft spectral components seen in some bright ULXs
may be interpreted as an accretion disk spectrum modified by a jet
that is responsible for the dominant power-law spectral component.

\begin{figure}
\centerline{
 \includegraphics[width=6.1truecm]{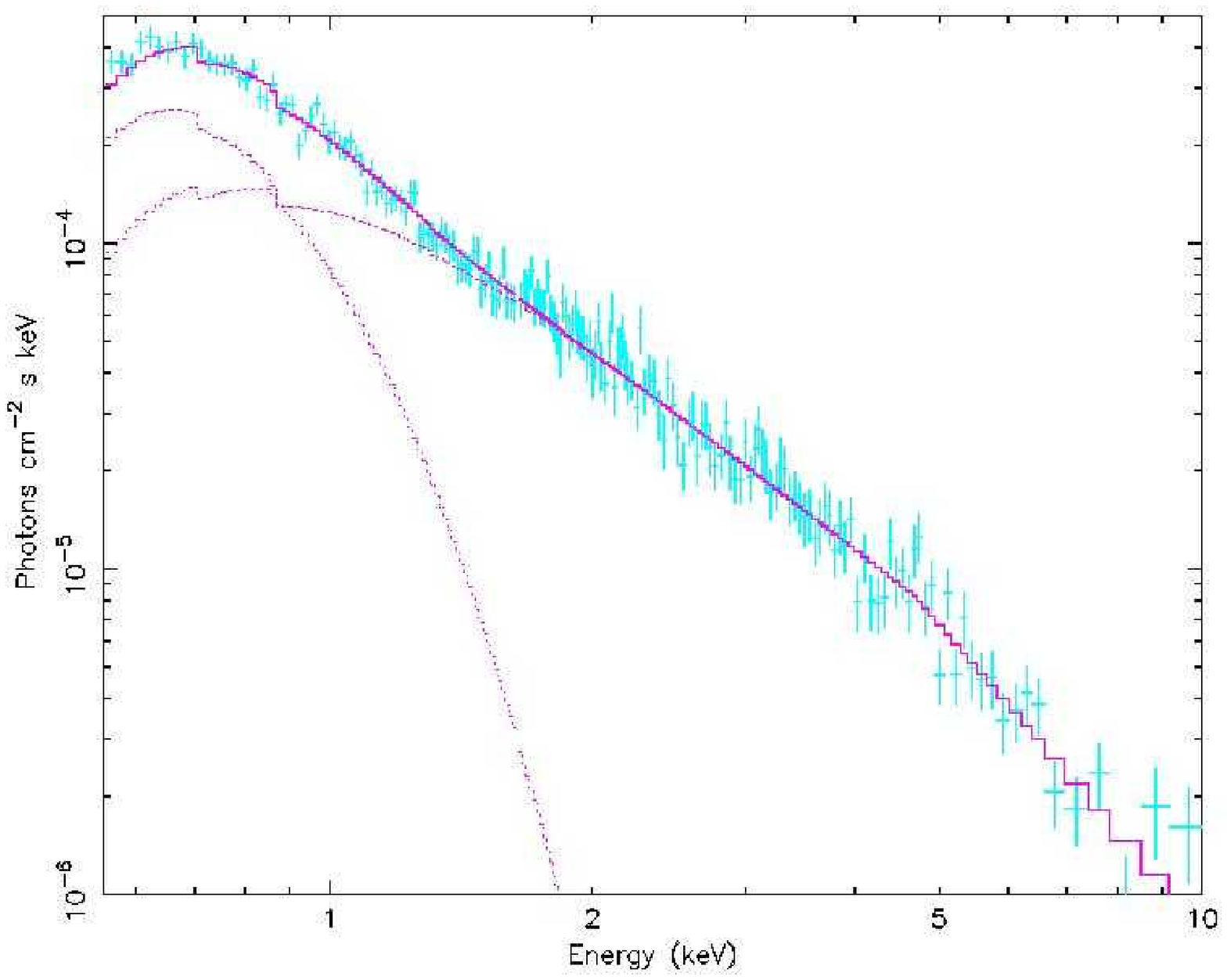}
\hspace{0.30truecm}
 \includegraphics[width=5.9truecm]{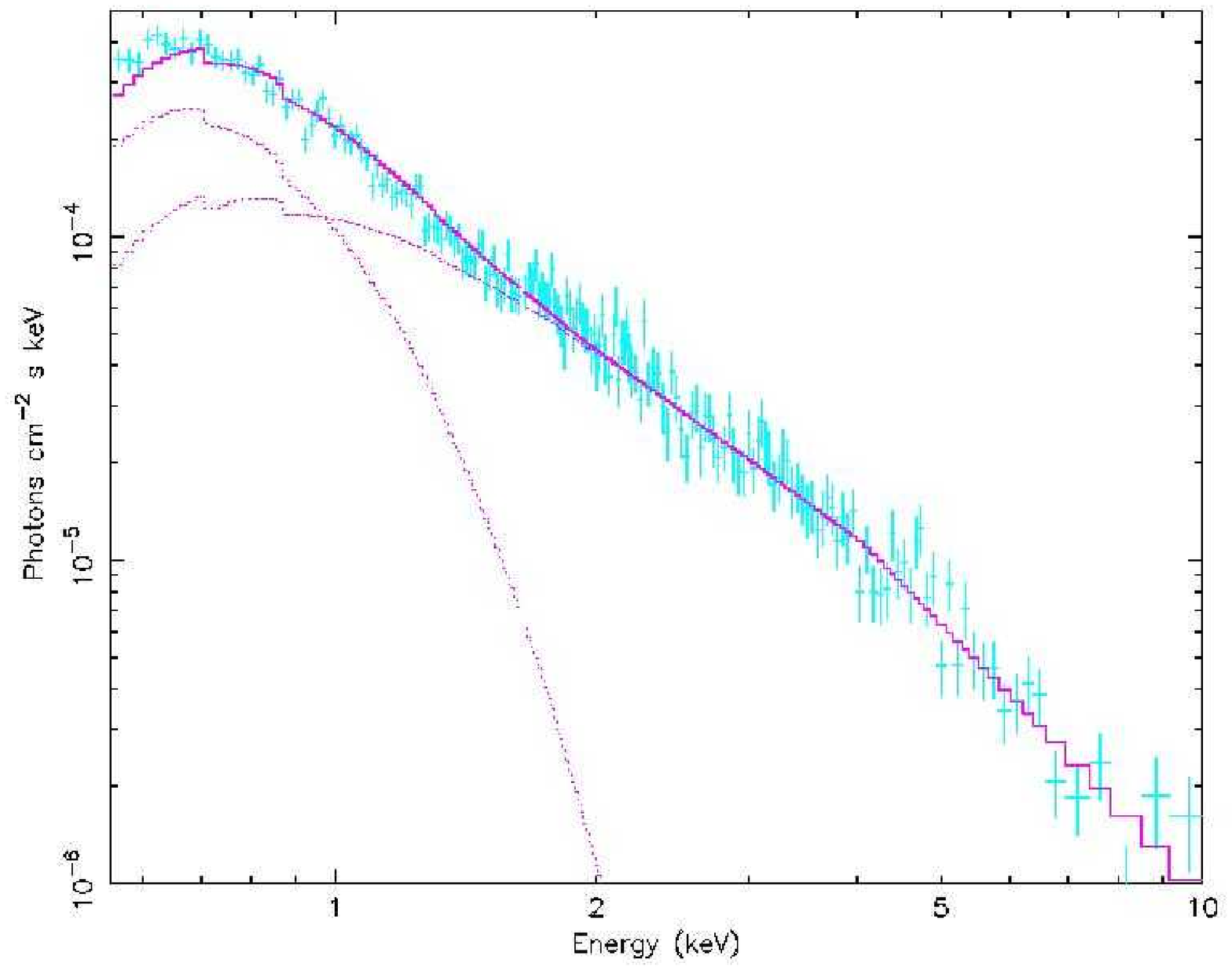}
}
\centerline{
 \includegraphics[width=6.4truecm]{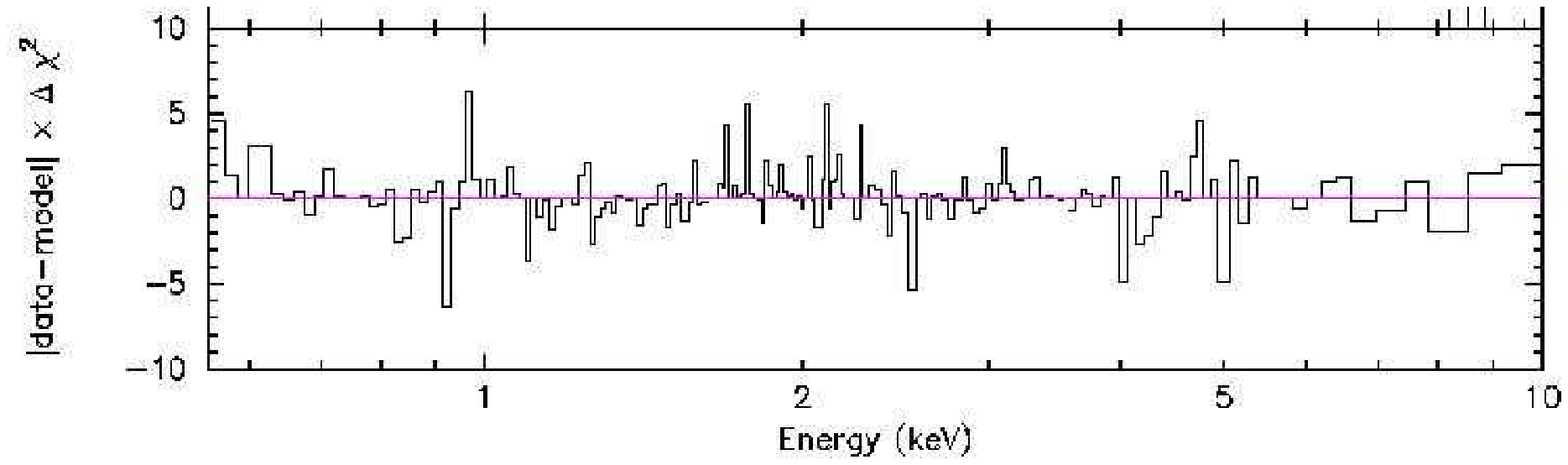}
 \includegraphics[width=6.3truecm]{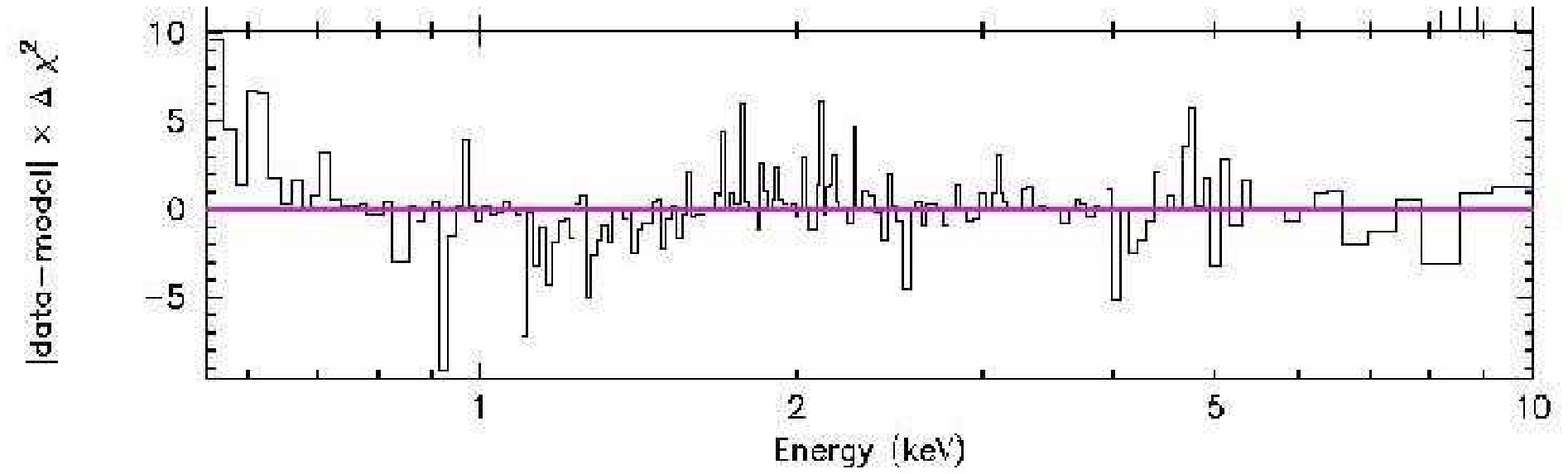}
}
\caption{\textit{XMM-Newton}/EPIC unfolded spectra of
  ULX NGC\,4559 X-7. The
  solid line is the total model spectrum. The dotted lines are the
  broken power-law plus MCD models (left) and the broken power-law plus OMMCD models
  (right). $\chi^2$ residuals are shown underneath. Best fit model parameters are
  shown in Table~\ref{tab:x1}.}
  \label{fig:x1}
\end{figure}

Figure~\ref{fig:x1} shows the \textit{XMM-Newton}/EPIC
spectrum of ULX X-7 in NGC\,4559. The spectrum of this bright ULX cannot adequately be
fitted with a single power-law model. A broken power-law plus MCD 
model provides an acceptable fit. The low inner disk temperature
required by the MCD fit to the soft component implies a black hole mass $\Mbh \sim
1.4 \times 10^3 M_\odot$. We also show in Fig.~1 an
almost identical spectral fit to the data using a similar broken
power-law plus an OMMCD model. The best fit model parameters are listed in
Table~\ref{tab:x1}. The OMMCD parameters correspond to a $\Mbh \approx 155\,M_\odot$ black
hole accreting at $\dot M_{\rm a} \approx 3 \dot M_{\rm Edd}$, but with only
$L_{\rm d} \approx 0.3 L_{\rm Edd}$ being emitted by the disk. According to
this model, the bulk of the accretion power is removed from the disk
and injected into a jet, which is responsible for the hard power-law
spectral component. However, only a small fraction  of the
jet power is dissipated in the form of radiation, since jets are
highly inefficient emitters. Relativistic beaming is then responsible for
boosting the X-ray emission to the high observed levels.

\subsection{Jet emission}

We calculated the jet spectral energy distribution for two different
relativistic beaming scenarios (\cite[K\"{o}rding, Falcke \& Markoff
  2002]{Kording02}) : 1. the microblazar scenario, where the jet is pointing
close to our line-of-sight and the observed emission is thus strongly
beamed; and 2. the microquasar scenario, where the jet is directed at
larger angles and hence, there is less contribution from relativistic
beaming to the observed X-ray emission. We used a simple radiative transfer
model to take into account the synchrotron optical depth. The details
of the microblazar and microquasar spectral models can be found in
\cite[Freeland \etal\ (2006)]{Freeland06}.
\begin{table}
  \begin{center}
  \caption{Best fit parameters for the spectral models used in Figure~1.}
  \label{tab:x1}
  \begin{tabular}{lc}\hline
  \multicolumn{2}{c}{broken power-law+MCD model} \\ \hline
  $\NH (\times 10^{21} \, {\rm cm}^{-2})$ & $2.3$ \\
  $kT_{\rm in} \, ({\rm keV})$ & $0.16$ \\
  normalization & $162.2$ \\
  $\Gamma_1$ & $2.11$ \\
  $E_{\rm break} (\rm {keV})$ & $4.66$ \\
  $\Gamma_2$ & $3.11$ \\
  normalization ($\times 10^{-4}$) & $2.2$ \\
  $\chi^2 / {\rm dof}$ & $202.6/195$ \\
  \hline
  \end{tabular}
\hspace{1.0truecm}
  \begin{tabular}{lc}\hline
  \multicolumn{2}{c}{broken power-law+OMMCD model} \\ \hline
  $\NH (\times 10^{21} \, {\rm cm}^{-2})$ & $2.3$ \\
  $\Mbh (M_\odot)$ & $ 155$ \\
  $\Mdota (\times 10^{-5} M_\odot \, {\rm yr}^{-1})$ & $1.3$ \\
  $L_{\rm d}/L_{\rm Edd}$ & $0.3$ \\
  normalization ($\times 10^{-6}$) & $1.0$ \\
  $\Gamma_1$ & $2.04$ \\
  $E_{\rm break} (\rm {keV})$ & $4.05$ \\
  $\Gamma_2$ & $2.79$ \\
  normalization ($\times 10^{-4}$) & $1.99$ \\
  $\chi^2 / {\rm dof}$ & $264.7/196$ \\
  \hline
  \end{tabular}
  \end{center}
\end{table}

\begin{figure}
\centerline{
 \includegraphics[width=6.0truecm]{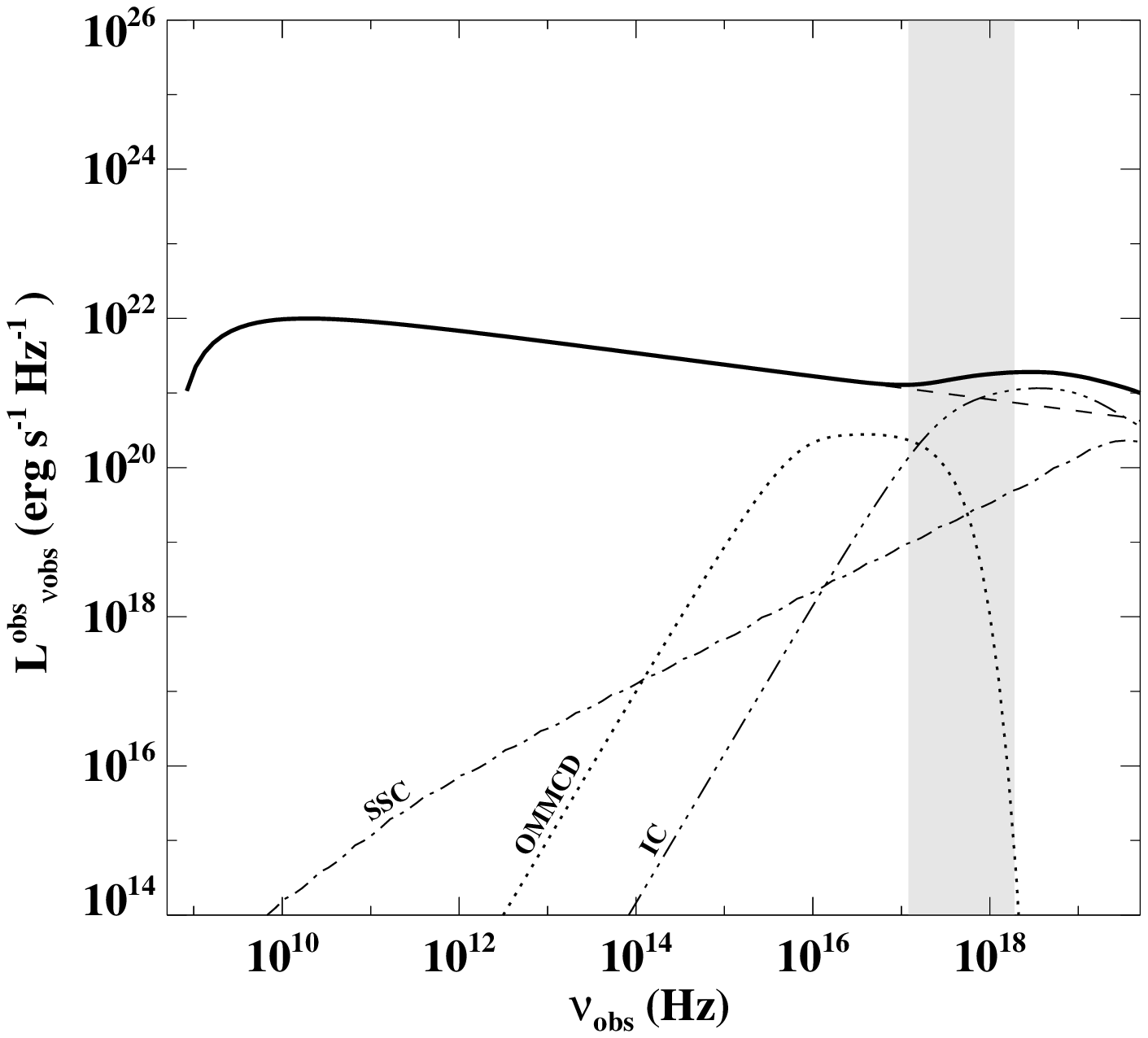}
\hspace{0.25truecm}
 \includegraphics[width=6.0truecm]{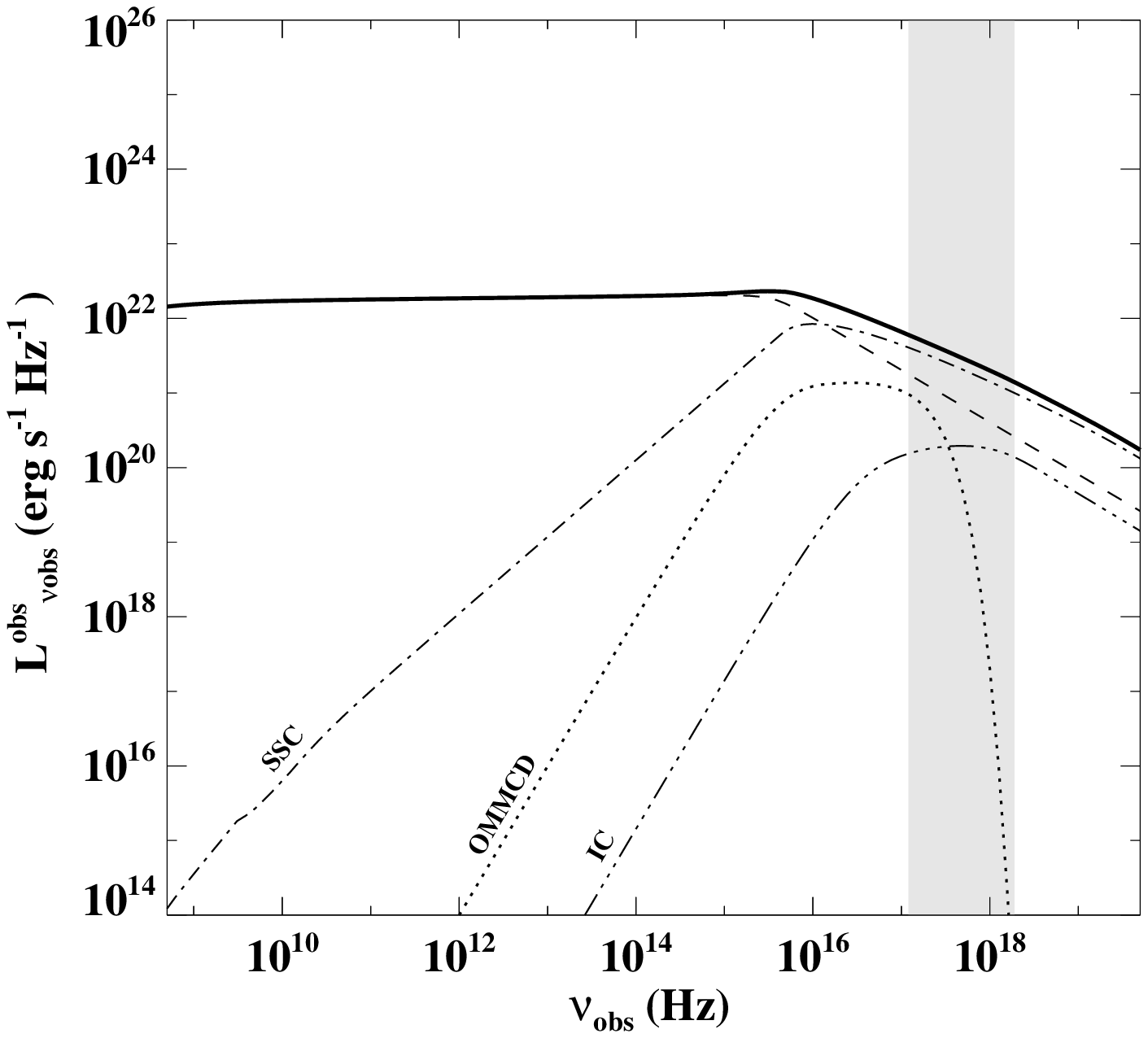}
}
\caption{The predicted spectral energy distributions for the
  microblazar (left) and microquasar (right) scenarios for ULXs. The
  microblazar model has $\Mbh = 5 M_\odot$, $\Mdota = \dot M_{\rm
    Edd}$, $\delta = 8.4$ and $\Lx \approx 3 \times 10^{39} {\rm erg
    \, s^{-1}}$. The
  microquasar model has $\Mbh = 20 M_\odot$, $\Mdota = \dot M_{\rm
    Edd}$, $\delta = 1.6$ and $\Lx \approx 5 \times 10^{39} {\rm erg
    \, s^{-1}}$. The shaded region indicates the $0.5 - 8\, {\rm keV}$
bandpass.}
  \label{fig:Lnu}
\end{figure}

The spectral modelling results shown in Figure~\ref{fig:Lnu} confirm
that both the microblazar and microquasar scenarios can produce X-ray
luminosites sufficiently high to be consistent with ULXs. The
microblazar model, however, predicts strong deviations from a
power-law in the $0.2 - 10\,{\rm  keV}$ bandpass resulting from
strongly beamed nonthermal
Comptonization. This is inconsistent with the observational data to
date. The microquasar model, on the other hand, predicts an
approximately power-law hard X-ray spectrum.

Both models predict similar radio properties. The synchrotron radio
spectra are approximately flat at $5\,{\rm GHz}$ and the specific
radio power is $L_\nu \approx 10^{22} {\rm erg \, s^{-1}
  Hz^{-1}}$. This corresponds to a flux density $S_\nu \approx 10 \,
\mu \, {\rm Jy}$ at at distance of $1\,{\rm Mpc}$. This is more than
two orders of magnitude below the levels measured for the few cases
where radio sources have been found associated with ULXs. Our
theoretical results support other pieces of evidence ruling out
beamed jet emission as the origin of ULX radio counterparts. Our
results are also consistent with the overwhelming excess of non-detections over
detections found in the deepest ULX radio counterpart search to date,
down to detection limits $\approx 60 \, \mu {\rm Jy}$ with the VLA
(\cite[K\"{o}rding, Colbert \& Falcke 2005]{KordColFalck05}). Note
that a typical Galactic XRB placed at a distance of $1\,{\rm Mpc}$
would have a flux density of only $\simeq 1\,\mu{\rm Jy}$.  

\section{Summary}

We have argued that ULXs appear to be in a persistently high/hard spectral
state and that they may represent an extreme sub-population
of XRBs. According to this
interpretation, ULXs must be accreting at very high rates (at least
Eddington) and must possess a strong corona and/or jet. We have outlined
a theoretical model that satisfies these criteria and that explicitly
identifies the disk-jet coupling mechanism as a magnetic
torque. The model predicts that ULXs should possess an accretion disk
that is substantially cooler at small radii than a standard disk at
the same $\Mdota$. This offers a viable explanation for the
unusually soft component seen in the spectra of some bright ULXs. We
fitted the $\textit{XMM-Newton}$ spectrum for ULX X-7 in NGC\,4559 and
found that the modified disk fit to the soft component requires a
much lower black hole mass ($\Mbh \approx 155 M_\odot$) than a standard
disk fit ($\Mbh \simeq 1400 M_\odot$).
We also presented theoretical spectral modelling results showing
that relativistic beaming can account for the observed X-ray
luminosities of ULXs without resorting to extreme black hole
masses. If ULXs are indeed relativistically beamed, we predict they
should exhibit unresolved, flat-spectrum radio cores with fluxes $\ltapprox
0.01\,{\rm mJy}$. Such radio counterparts have not yet been detected.

\begin{acknowledgments}
ZK acknowledges a University of Sydney R\&D Grant. RS acknowledges
a NASA \textit{Chandra} grant.
\end{acknowledgments}

\end{document}